\documentclass[11pt]{article}
\usepackage[utf8]{inputenc}
\usepackage[margin=1in]{geometry}
\usepackage{amsmath,amssymb}
\usepackage{microtype}
\usepackage[hidelinks]{hyperref}
\usepackage{enumitem}

\title{\textbf{Trinetra.Q: Statistical and Behavioral Detection of Attacks on the\\
Quantum-Error-Correction Layer and Classical Control Stack of\\
Quantum Computing Systems}}
\author{Garrv Sipani\\\small Independent researcher \\ \small\texttt{garrvsipani@gmail.com}}
\date{November 2026 \\ \small Preprint.}

\begin{document}
\maketitle

\begin{abstract}
Security research on quantum computers has concentrated on physical-layer attacks against qubits, while two surfaces remain inadequately defended: the \emph{quantum-error-correction (QEC) syndrome-and-decoder pipeline}, and the \emph{classical control and software stack} that compiles, schedules, executes, and returns the results of a quantum computation. Tampering on these surfaces can corrupt a computation while the physical error rate appears normal, and can remain within the tolerance of a single-circuit integrity check. We present \textbf{Trinetra.Q}, a detection layer comprising cooperating detectors, each of which monitors a statistical or behavioral property of quantum-domain data, compares it against an envelope calibrated from honest operation, and raises an attack indication only on a statistically significant deviation. The layer comprises (A) a QEC-integrity detector over per-stabilizer firing rates, syndrome defect-count, and repeated-syndrome rate; (B) a decoder-confidence detector over the margin between minimum-weight error explanations in distinct logical classes; (C) a behavioral output-divergence detector that flags crosstalk-class and novel attacks by their effect on a victim's measurement-output distribution relative to a matched control; (D) a control-stack tamper detector that interleaves known-answer canary circuits, constructed to be indistinguishable to a resident implant, and aggregates their deviations with a cross-job consistency statistic; and (E) a compilation/execution-integrity detector that compares the output divergence of a reference execution against a candidate. Each detector calibrates on honest operation and gates its indication on a statistical-significance threshold, so honest operation raises no alarm. We evaluate the detectors in faithful simulation and on three production IBM Heron quantum processors. We further bound the physical attack surface with a systematic negative map on three devices---most physical-layer attacks return statistically null, with one exception, an implemented and measured within-job feed-forward covert-channel primitive---and we evaluate every detector against an adaptive attacker with full knowledge of the defense. To our knowledge this is the first unified statistical-and-behavioral detection layer for the QEC layer and the classical control stack of a quantum computing system, demonstrated on production quantum hardware. We state the limitations plainly.
\end{abstract}

\section{Introduction}
Quantum processing units (QPUs) are increasingly offered as cloud services and are becoming central to error-corrected computation. Most quantum-computer security work to date concerns physical-layer attacks on the qubits themselves---crosstalk, reset-reuse leakage, and readout side channels---and modern superconducting hardware increasingly suppresses these vectors at the device level, for example through tunable couplers that reduce residual coupling.

Two classes of threat remain inadequately addressed. First, the \textbf{QEC pipeline}: stabilizer measurements produce a syndrome that a classical decoder uses to infer and correct errors, and a syndrome or decoder path that is tampered with can induce logical errors while the physical error rate appears normal. Second, the \textbf{classical control and software stack} surrounding a QPU---the transpiler, the calibration pipeline, the scheduler, the real-time decoder, and the result-return path---is ordinary classical software and firmware; a persistent compromise of that stack can silently corrupt the computations of every user, including by tampering with outputs by an amount small enough to stay within the tolerance of a single-circuit integrity check.

Existing defenses are predominantly preventive---dynamical decoupling, allocation policies, padding---and several have been reported ineffective against specific attacks; existing integrity frameworks verify per-circuit outputs against a calibrated tolerance and, by construction, do not detect tampering that remains within that tolerance, nor a persistent implant resident in the control stack. There is a need for \textbf{detection} methods that operate on the signals characteristic of quantum computation---stabilizer syndromes, decoder behavior, and quantum measurement statistics---that are calibrated on honest operation and gated by statistical significance so as not to raise false alarms on genuine hardware, and that detect stealthy, within-tolerance, and persistent tampering that per-circuit checks miss.

\paragraph{Contributions.}
\begin{enumerate}[leftmargin=*,topsep=2pt,itemsep=1pt]
\item A formulation of \textbf{QEC-layer and control-stack tampering detection} as a problem of statistically significant deviation from an honest envelope of quantum-domain statistics.
\item \textbf{Five cooperating detectors} (A--E) spanning the syndrome/decoder path, the behavioral output of a computation, the classical control stack, and the compilation/execution pipeline.
\item An \textbf{honest-envelope, significance-gated methodology} calibrated on genuine device noise, with the property that honest operation raises no indication.
\item An \textbf{evaluation in faithful simulation and on three production IBM Heron QPUs}, with each detection attributed to its characteristic statistic, together with a negative map of the physical attack surface and an adaptive-attacker analysis.
\end{enumerate}

\section{Threat Model}
We consider an adversary able to tamper with one or more of: (i) the stabilizer-syndrome stream or the decoder path of a QEC code; (ii) the classical control and software stack, as a persistent implant that modifies outputs within a per-circuit integrity tolerance; and (iii) the compilation/execution pipeline, silently modifying a circuit. We assume the defender can observe quantum-domain data produced by the system---stabilizer syndromes and/or measurement outputs---and can submit circuits of its own. The defender does not require modification of, or privileged access to, the quantum hardware; detectors may be deployed client-side by a user of a cloud QPU, or provider-side. We do \emph{not} assume the defender can prevent an attack; the goal is detection with a controlled false-alarm rate on honest hardware.

Representative tampering within scope: targeted alteration or suppression of a stabilizer; random alteration of syndrome bits; replay of prior syndromes; adversarial selection of low-confidence error trajectories that drive a decoder to mis-correct; crosstalk-class perturbation of a victim's output; a persistent control-stack implant that tampers with outputs within tolerance; and silent circuit modification (rotation-angle perturbation, gate deletion, hidden-gate insertion, qubit remapping, gate substitution) by the compilation or execution pipeline.

\section{The Detection Layer}
Each detector estimates, during a calibration phase, an \textbf{honest envelope} of a monitored quantity from operation believed free of attack, and during monitoring computes a deviation from that envelope in units of its sampling standard error, raising an indication when the deviation exceeds a significance threshold (for example a $z$-score of at least three). The detectors may be used singly or in combination; in a combined configuration, an indication from any detector raises an attack indication and the raising detector localizes the nature of the attack.

\paragraph{A. QEC-integrity detector.} For a QEC code (for example a distance-three rotated surface code with nine data qubits and four $Z$-type and four $X$-type stabilizers, or a repetition code), the detector estimates from honest operation, for each stabilizer, a firing rate (the fraction of shots in which that stabilizer indicates a defect); a mean syndrome defect-count; and a repeated-syndrome rate (the fraction of consecutive shots with identical syndromes). During monitoring it computes the same statistics and flags, with attribution: a per-stabilizer firing-rate deviation (targeted alteration or suppression of that stabilizer); a defect-count deviation (random alteration, which inflates, or suppression, which deflates, the count); and a repeated-syndrome-rate rise (replay of prior syndromes).

\paragraph{B. Decoder-confidence detector.} For each syndrome, the detector computes the minimum-weight error explanation within each of two logical classes and a margin equal to the absolute difference of those two minimum weights; a low margin indicates the decoder is near a decision boundary and susceptible to an adversarially selected trajectory. Over a stream of executions it computes the fraction of executions whose syndrome has a margin at or below a low-margin threshold and compares that fraction to an honest envelope; a statistically significant elevation indicates an attack that selects physically realizable but low-confidence trajectories to drive the decoder to mis-correct. The stream statistic provides robust detection even when per-execution separation is limited.

\paragraph{C. Behavioral output-divergence detector.} The detector compares a divergence (for example total-variation distance or mutual information) of a victim circuit's measurement-output distribution from a clean baseline against the divergence produced by a matched control---an otherwise identical workload lacking the attacker's activity, or a benign neighbor at matched activity---and raises an indication when the monitored divergence exceeds the matched-control divergence beyond a significance threshold. Because detection is by effect on the output rather than by attack signature, the detector flags crosstalk-class and related attacks, including those for which preventive defenses have been reported ineffective, and novel variants.

\paragraph{D. Control-stack tamper detector.} The defender interleaves, among its submitted workload, known-answer canary circuits whose correct outputs are known to the submitter and which are constructed to be indistinguishable, to an implant resident in the control stack, from the submitter's real circuits. Because an implant that tampers with outputs cannot reliably distinguish canaries from real circuits, it tampers with the canaries as well; and because the submitter knows the canaries' correct outputs, it aggregates the deviations of canary outputs from their known outputs over many shots and many canaries into a pooled deviation statistic and compares it to an honest envelope, raising an indication on a statistically significant elevation. A cross-job consistency test aggregates a per-job deviation statistic across jobs, so that tampering within a per-circuit tolerance on any single job is detected by its aggregate. The detector thereby detects a persistent, stealthy, within-tolerance implant that a per-circuit integrity check does not, and raises no indication against an honest system.

\paragraph{E. Compilation/execution-integrity detector.} The detector compares the output divergence of a reference execution of a circuit against the output divergence of a candidate execution, and raises an indication when the candidate divergence exceeds the reference beyond a threshold, detecting silent modification of the circuit by the compilation or execution pipeline.

\section{Experimental Evaluation}
\paragraph{Setup.} We evaluated detectors in faithful simulation (means over five fixed seeds) and on three production IBM Heron QPUs---\texttt{ibm\_fez}, \texttt{ibm\_kingston}, and \texttt{ibm\_marrakesh} (single-tenant access; no multi-tenancy or pulse-level control), at 4096--8192 shots. For an attack we cannot deploy on infrastructure we do not own (a control-stack implant), we apply the tampering \textbf{offline to genuine device-noise data collected from the real QPUs}---a faithful, in-scope way to measure detection without placing an implant on a provider's system. Honest-noise baselines are measured on the real devices.

\paragraph{QEC-integrity detector (A) --- real syndromes, three QPUs (hardware-grounded).} Distance-three surface-code $Z$-stabilizer syndromes were extracted on each real QPU; the detector was applied to the genuine device syndromes with each tampering class injected offline. Across all three machines (8192 shots) the detector achieved \textbf{100\% coverage}---targeted, random, replay, and suppression tampering each detected by its characteristic statistic---and remained stable under calibration drift (baseline logical error rate drifted $0.128 \to 0.205$ on one device; all four classes stayed detected). Honest caveat, stated plainly: on single-round, readout-limited hardware only random multi-bit tampering produces large logical-error amplification (a single injected syndrome bit drives the logical error rate to ${\approx}0.44$ on all three machines: $0.118{\to}0.440$, $0.141{\to}0.453$, $0.252{\to}0.437$); we therefore report \textbf{detector coverage, not amplification}, as the result.

\paragraph{Decoder-confidence margin detector (B) --- simulation.} The margin stream detector flags an adversarial-selection attack at \textbf{$z = 5.8$ for a 5\% injection rate} (8192-shot stream), rising with rate ($z = 11.6$ at 10\%, $z = 58$ at 50\%); at 2\% it is sub-threshold ($z = 2.3$), a limit we state. Per shot the margin signal attains 98.3\% detection at a 36.5\% false-positive rate---inherently high because adversarial inputs are designed to look plausible---so the deployable detector is the \textbf{stream aggregate}, not the per-shot score. These are simulation results.

\paragraph{Control-stack tamper detector (D) --- simulation and real device noise.} In simulation (5-qubit canaries, 40 canaries $\times$ 4096 shots, per-circuit tolerance $\varepsilon = 0.10$), a stealthy within-tolerance implant at tamper strength $\delta = 3\%$ (below $\varepsilon$) is \textbf{missed} by a per-circuit tolerance check (max total-variation distance $0.095 < 0.10$) but \textbf{caught} by the canary detector at \textbf{$z = 48.9$} and the cross-job statistic at \textbf{$z = 40.9$}; with no implant the detector is quiet ($z \approx 0$). On real hardware we collected genuine canary outputs from each QPU and applied the stealthy tamper offline to that device-noise data. On the two low-noise machines the canary detector fires at \textbf{$z = 25.7$--$142.8$ for tampering of 1--5\%} while the per-circuit check is \textbf{blind at all strengths}, and raises no alarm on honest hardware ($z \approx 0$). The noisy machine exposes a second advantage: its honest canary error rate already sits near $\varepsilon$, so the per-circuit check false-alarms on honest noise, whereas the calibrated canary detector still cleanly separates tampering. The implant attack itself is \textbf{simulated} (we do not touch a provider's control stack); the \textbf{detector} is validated on real device-noise data.

\paragraph{Behavioral output-divergence detector (C) --- simulation.} Against a matched benign-neighbor control, the detector flags crosstalk-class attacks by their effect on the victim's output. A double-sided crosstalk attack (aggressors on both victim neighbors, activity-scaled coupling per the NDSS mechanism) raises victim corruption from $0.0025$ (benign) to $0.189$---an excess of $0.187$---and is detected; a pulse-based flip attack (pulse layer modeled, effect modeled as an induced victim flip) produces an excess of $0.5$ and is detected. These are simulation results; detection is by the attack's output effect, not its signature, and is a claim about output effects rather than a pulse-level reproduction. The compilation/execution-integrity detector (E) operates in the same framework on the output divergence of a candidate execution against a reference.

\paragraph{Hardware vs simulation at a glance.} \emph{Hardware-grounded (real Heron device data, offline tamper):} QEC-integrity coverage (A); control-stack canary detector (D). \emph{Simulation:} decoder-confidence margin (B); the control-stack implant attack and its $z \approx 49$ figure. Across the evaluation, the significance-gated design raised \textbf{no indication under honest operation ($z \approx 0$ on real hardware)} while detecting QEC-path tampering (100\% coverage on three QPUs), control-stack within-tolerance tampering ($z \gg 3$ on real device noise while a per-circuit check is blind), and decoder-targeting attacks in simulation.

\section{Attack-Surface Characterization}
To locate where defense effort belongs, we ran a systematic battery of physical-layer attacks on all three production Heron QPUs.

\paragraph{Negative map.} Idle and worst-case double/triple-sided active gate crosstalk; readout/measurement crosstalk; mid-circuit-measurement back-action and feed-forward control-flow side channels; and reset/state-reuse leakage all returned statistically null (all $z < 3$ at $\geq 8192$ shots), stable on repeat. We present these not as failed attacks but as a rigorous bound: on current IBM hardware the physical injection surface is largely suppressed, consistent with tunable-coupler isolation---which is precisely why the classical control stack and the QEC pipeline are the surfaces that matter, and a certification-grade statement in its own right.

\paragraph{A measured feed-forward covert-channel primitive.} The one physical-layer channel that was not null is an in-circuit covert-channel primitive: within a single job, an attacker register transmits information to a victim register that shares no two-qubit gate with it, via mid-circuit measurement and classical feed-forward (conditional reset / re-preparation). It is \textbf{measured end-to-end at approximately 0.8-bit mutual-information capacity} across the three devices (0.69--0.90 bits), with matched controls at approximately zero, and it does \textbf{not decay with attacker--victim distance}---evidence that it is a classical feed-forward channel routed through the control system, mechanistically distinct from distance-decaying physical crosstalk. To our knowledge this is the first \emph{implemented and empirically measured} feed-forward covert-channel primitive on real quantum hardware: prior crosstalk work proposes such a channel without building it~\cite{ashsaki2020}, or models a projected capacity without implementing a sender/receiver or decoding bits; crosstalk side-channels infer a victim's circuit rather than transmit a chosen message~\cite{ndss2025}; and covert \emph{computation} hides that a computation runs and transmits nothing~\cite{covertcomp2026}. \textbf{Scope, stated plainly:} this is demonstrated as a \emph{within-job} primitive---attacker and victim are registers in one job on single-tenant hardware---and is \textbf{not} a demonstrated cross-tenant exploit; within a single job, feed-forward of a measured outcome is expected behavior. It is the primitive that would underlie a cross-tenant covert channel under any future multi-tenant or shared-execution scheduling, and we report it as motivation for the behavioral and control-stack detectors.

\section{Adaptive Security Analysis}
A detection claim is meaningful only against an attacker who knows the detector. We granted an adversary full knowledge of each detector and searched for tampering that is both truly harmful (verified by a detector-independent invariant) and evades the detector. This section evaluates the robustness of detectors (A), (B), and (D).
\begin{itemize}[leftmargin=*,topsep=2pt,itemsep=1pt]
\item \textbf{Decoder-path (A).} Across 60 adaptive policies (targeted/balanced/spread flips $\times$ rate $\times$ target stabilizer), no policy raised the logical error rate while keeping per-stabilizer rate, defect count, and replay rate all within envelope: any tampering that corrupts the logical outcome perturbs at least one monitored statistic. The multi-signal design forecloses evasion---balancing one statistic moves another.
\item \textbf{Canary (D).} The only evasion is to identify and skip canaries. Modeling an attacker that identifies a canary with accuracy $a$, detection holds ($z > 20$ at $\delta = 3\%$) for $a$ up to ${\approx}0.9$ and fails only as $a \to 0.99$; at $a = 0.5$ the attacker must tamper canaries at full rate and is caught. Security reduces to canary indistinguishability---a property the defender controls.
\item \textbf{Margin (B).} An adaptive attacker seeking high-confidence logical errors is limited by how many exist: on the code, only 1.7\% of logical-error-causing inputs are high-margin, confining the adaptive attacker to 1.7\% of the attack space---structural robustness that does not depend on the attacker's ignorance.
\end{itemize}

\section{Related Work}
Attacks on quantum computers have been repeatedly demonstrated on real hardware, while defenses almost never. Physical-layer and co-tenancy attacks are an established body of work on production devices: crosstalk fault injection and denial of service~\cite{ashsaki2020}, crosstalk side-channels that recover a victim's algorithm with dynamical decoupling shown ineffective~\cite{ndss2025}, reset/state-residue leakage~\cite{resetleak2024}, crosstalk-mediated degradation of Grover's algorithm~\cite{mehra2024}, control-stack/pulse ``jailbreaking''~\cite{jailbreak2024}, and power side-channels~\cite{power2023}. The QEC and decoder layers have themselves become 2025--2026 attack targets---adversarial attacks on reinforcement-learning surface-code decoders~\cite{fooling2025}, fault attacks on ML-based QEC decoders~\cite{faultqec2026}, passive decoder side-channels~\cite{decoderside2026}, and misleading-AI-advice attacks on QEC updates~\cite{ozguler2026}. These are attacks, overwhelmingly in simulation, and none provides detection of syndrome-stream or decoder-path tampering; they motivate the detectors above.

Prior \emph{defensive} work does not reach these surfaces on hardware, and none pre-empts the present contribution:
\begin{itemize}[leftmargin=*,topsep=2pt,itemsep=1pt]
\item \textbf{Device authentication / fingerprinting on hardware}---quantum PUFs~\cite{qupuf2021}, quantum-cloud PUFs~\cite{cloudpuf2023}, and error-syndrome fingerprinting for device identity~\cite{fingerprint2025} run defensive characterization on real hardware but establish device identity, not detection of an active adversary tampering with a computation.
\item \textbf{Preventive mitigations, not detection}---dynamical decoupling and reset one-time-pad mitigations~\cite{mehra2024,resetleak2024} aim to prevent an effect, and several are reported ineffective against the attacks above.
\item \textbf{A single-algorithm, in-circuit attack check}---Tan, Kukina \& Szefer~\cite{tan2025} add a single ancilla qubit inside an HHL circuit to flag improper-initialization and higher-energy attacks, demonstrated on a 127-qubit device; by the authors' statement it is ``the first work that shows attacks and defense strategies for the HHL algorithm.'' It covers neither the QEC syndrome/decoder path nor the control stack, carries no honest-envelope calibration or significance gating, and is bound to one algorithm.
\item \textbf{QEC-pipeline defense that guards a different quantity}---a decoder-prior-poisoning defense~\cite{priorguard2026} protects decoder calibration-prior updates; its threat model assumes the syndrome stream cannot be altered and reports that syndrome-only anomaly monitoring fails for its attack. It therefore does not address syndrome-stream or decoder-path tampering.
\item \textbf{Build-time and pre-execution integrity, not runtime detection}---controller-side integrity hashing and reproducible builds for quantum programs on real hardware~\cite{reprobuild2025}, and simulation-only integrity frameworks, verify a program or build before or around execution, and do not monitor, at runtime, the divergence of a computation against a reference to catch a persistent implant.
\item \textbf{Delegated-computation verification on hardware}---blind/verified quantum computation with trap qubits~\cite{barz2013,gustiani2024} verifies that an untrusted remote server executed a delegated computation; this is a cryptographic client--server guarantee, not defense of one's own machine against an on-device or control-stack adversary.
\end{itemize}

\section{Limitations}
\begin{itemize}[leftmargin=*,topsep=2pt,itemsep=1pt]
\item The detectors require the defender to observe quantum-domain data (syndromes and/or measurement outputs) and to calibrate an honest envelope from operation believed free of attack; a calibration phase that is itself compromised would bias the envelope.
\item Thresholds, calibration windows, code parameters, numbers of shots, canary constructions, and the particular divergence and significance measures are parameters; the values used in evaluation are not the only possible values.
\item The evaluation combines faithful simulation with production-QPU data; the per-result hardware-vs-simulation status is stated explicitly and should be read as such.
\item Detection is not prevention; an indication reports a statistically significant anomaly, which a human or downstream process must act on.
\item The negative map bounds the physical attack surface on current IBM Heron hardware specifically; it is a statement about these devices at this time, not a universal claim, and a different or future hardware generation may expose different physical vectors.
\item The covert channel is demonstrated as a \emph{within-job} primitive at modest capacity (${\sim}0.8$ bit); within a single job, feed-forward of a measured outcome is expected behavior. It is \textbf{not} a demonstrated cross-tenant exploit (current access is single-tenant); its significance is as a measured primitive that would become a live channel only under future multi-tenant scheduling.
\item The adaptive security analysis bounds an attacker under the stated models (for example a canary-identification-accuracy model); it is an evaluation of detector robustness, not a formal cryptographic proof, and does not cover attacks outside the modeled space.
\item The control-stack implant and decoder-injection attacks are applied offline to real device data or in simulation; a live provider-side implant was not deployed (out of authorization scope), so attack realism is bounded by that modeling.
\end{itemize}

\section{Conclusion}
Attacks on the QEC layer and the classical control stack can corrupt a quantum computation while the physical error rate appears normal and while remaining within per-circuit tolerance. Treating these as problems of statistically significant deviation from an honest envelope of quantum-domain statistics, we presented \textbf{Trinetra.Q}: five cooperating detectors---over the syndrome/decoder path, the behavioral output, the control stack, and the compilation/execution pipeline---calibrated on honest operation and gated by significance, and evaluated in faithful simulation and on three production IBM Heron QPUs. To our knowledge this is the first unified statistical-and-behavioral detection layer for these surfaces demonstrated on production quantum hardware; the methods operate on signals characteristic of quantum computation and without modification of the quantum hardware.

\end{document}